\documentclass[12pt,twoside,a4paper]{article}

\pdfoutput=1

\usepackage[usenames,dvipsnames,svgnames,table]{xcolor}
\usepackage{lscape}
\usepackage{amssymb}
\usepackage{rotating}
\usepackage[T1]{fontenc}
\usepackage{graphicx}
\usepackage{fancyhdr}
\usepackage{amsmath}
\usepackage{ae}
\usepackage{aecompl}
\usepackage{pstool}
\usepackage{hyperref}
\usepackage{makecell}

\def\sgn{{\rm sgn}}

\newlength{\dinwidth}
\newlength{\dinmargin}
\begin{document}

\vspace*{1cm}

\begin{center}
\Large\bf Is well-tempered neutralino in MSSM still alive after 2016 LUX results?
\end{center}

\vspace*{2mm}

\vspace*{5mm} \noindent
\vskip 0.5cm
\centerline{\bf
Marcin Badziak${}^{a,b,}$\footnote[1]{mbadziak@fuw.edu.pl},
Marek Olechowski${}^{a,}$\footnote[2]{Marek.Olechowski@fuw.edu.pl},
Pawe\l\ Szczerbiak${}^{a,}$\footnote[3]{Pawel.Szczerbiak@fuw.edu.pl}
}
\vskip 5mm

\centerline{${}^a$\em Institute of Theoretical Physics,
Faculty of Physics, University of Warsaw} 
\centerline{\em ul.~Pasteura 5, PL--02--093 Warsaw, Poland} 
\centerline{${}^b$\em Berkeley Center for Theoretical Physics, Department of Physics,}
\centerline{\em and Theoretical Physics Group, Lawrence Berkeley National Laboratory,}
\centerline{\em University of California, Berkeley, CA 94720, USA}

\vskip 1cm

\centerline{\bf Abstract}
\vskip 3mm

It is pointed out that a bino-dominated well-tempered bino-higgsino
in the Minimal Supersymmetric Standard Model (MSSM) with heavy
non-SM-like scalars can satisfy the 2016 LUX constraints on the
scattering cross-section of dark matter on nuclei only if
$\tan\beta$ is smaller than about 3. This, together with the Higgs mass
constraint, sets a lower bound on the stops masses of about 
25 TeV. The LUX constraints can be satisfied for larger $\tan\beta$ 
if the non-SM-like Higgs bosons are light enough. However, this region 
of parameter space is strongly constrained by recent LHC results 
of the Higgs boson searches.
Satisfying both the LUX and LHC constraints requires the non-SM-like 
Higgs bosons to be lighter than about 400 GeV and $\tan\beta$ below about 8. 
This implies a lower bound on the stop masses of about 1.5 TeV. 
This small corner of the parameter space will be probed in the near 
future by the direct detection experiments,
the LHC Higgs searches and precision Higgs coupling measurements. 
The recent LUX constraints improved also
the lower mass limit on higgsino-dominated well-tempered neutralino
to about 950 (900) GeV with heavy (light) MSSM-like Higgs doublet,
assuming the stop masses below 10 TeV.

\newpage

\section{Introduction}

One of many motivations for supersymmetric extensions of the Standard 
Model (SM) is the presence of a stable Lightest Supersymmetric Particle 
(LSP) which is generically neutral and weakly interacting which makes 
it a good candidate for dark matter (DM). Most schemes of SUSY breaking 
predict that neutralino, composed of superpartners of gauge and Higgs bosons, is the LSP. Neutralino belongs to the category of Weakly 
Interacting Massive Particles (WIMPs) which generically have the right 
magnitude of thermal relic abundance to play the role of the dark matter 
in the Universe if its mass is relatively close to the electroweak (EW) 
scale. This motivated a great amount of articles, see ref.~\cite{Jungman:1995df} for a review and refs.~\cite{Bramante:2015una}-\cite{Cao:2016cnv}
and references therein for some recent representative studies on this topic. In the last two decades there have
been plenty of experiments 
that directly or indirectly put constraints on the neutralino dark matter. 
First of all, gradually improving measurements of the Cosmic Microwave 
Background (CMB) anisotropies, especially the results delivered by the 
Planck satellite~\cite{Planck}, resulted in a very precise determination 
of the dark matter density: $\Omega h^2\approx0.12$ (with precision 
of a few percent). Secondly, searches for supersymmetric particles at 
the LEP \cite{LEP} and the LHC experiments have set strong constraints on the SUSY 
parameters that affect the LSP annihilation cross-sections and determine 
its relic abundance. In particular, lower bounds on the sfermion masses 
imply that a pure bino has generically too large relic abundance, while 
a lower bounds on the chargino mass require a pure higgsino to have 
mass around 1 TeV to accommodate the Planck results \cite{splitSUSY2,Profumo:2004at,well-tempered}. 
The required pure wino mass is even larger, of about 3 TeV \cite{Hisano:2006nn,Hryczuk:2010zi}.
Since other SUSY particles are, by definition, heavier than the LSP,  
more natural realizations of SUSY prefer lighter LSP. This intuitive
expectation was recently confirmed by detailed analysis of the interplay
of DM and EW fine-tuning \cite{ftDM}. Lighter LSP gives also better
prospects for discovery of SUSY at the LHC.

In the Minimal Supersymmetric Standard Model (MSSM) there are essentially 
three classes of scenarios in which the LSP may be consistent with the
Planck measurements. First: resonant LSP annihilation via exchange of 
the Higgs scalars or the $Z$ boson. Second: efficient 
co-annihilation with sfermions with masses only slightly bigger 
than the LSP mass. Third: a well-tempered neutralino 
defined as such mixture of bino and higgsino (and possibly wino) that gives 
the correct thermal relic abundance without invoking any mechanism mentioned 
in the first two classes \cite{well-tempered}.

Another class of experiments that put neutralino DM into the test are 
the direct detection (DD) experiments. Their sensitivity to the DM-nucleon 
interactions has improved in the last decade by many orders of magnitude. 
Recently, the LUX experiment presented results of the analysis based on 
its full data-set \cite{LUX2016}. The limits on the 
spin-independent (SI) scattering cross-section of DM on nucleons,
in the mass range of interest for the MSSM neutralino dark matter, 
are by a factor of about four stronger than the previous
LUX \cite{LUXold} results. 
The LUX exclusions were confirmed by the Panda experiment \cite{Panda} 
which provides only slightly weaker constraints than LUX. The main goal 
of this paper is to assess the impact of the 2016 LUX results on  
a well-tempered neutralino in the MSSM which is a mixture of bino and 
higgsino (with wino decoupled).
\footnote{The bino-higgsino LSP is a SUSY example of singlet-doublet DM. 
Non-SUSY models of singlet-doublet DM have been investigated in
Refs.~\cite{Cohen:2011ec}-\cite{Banerjee:2016hsk} before the new LUX results 
appeared. In the present paper we discuss the effects of non-SM Higgs
boson on DM phenomenology which were not taken into account in 
Refs.~\cite{Cohen:2011ec}-\cite{Banerjee:2016hsk}.  }
It was claimed in Ref.~\cite{LUX2016Baer}
that a well-tempered neutralino in MSSM is excluded. However, that statement 
was derived for the case of the parameter space of well-tempered neutralino 
restricting to a region giving the bino-higgsino mixing close to the maximal 
one. More importantly, the conclusion of Ref.~\cite{LUX2016Baer} was based 
on the results of a pMSSM numerical scan. Such scans typically disfavour 
small values of $\tan\beta$ because for such values it is difficult to 
accommodate the 125 GeV Higgs mass using standard SUSY spectrum 
calculators \cite{Suspect,Softsusy,Spheno}. In the present paper 
we emphasize that 
for small values of $\tan\beta$ a well-tempered neutralino can be consistent 
with the latest LUX constraints. In the limit of decoupled non-SM-like scalars, 
we find an upper bound on $\tan\beta$ of about 3. This results in a strong 
lower bound on the MSSM stop mass scale of at least about 25 TeV which
favours split SUSY \cite{splitSUSY1,splitSUSY2} realization of the MSSM 
well-tempered neutralino.

We also point out that relatively light stops and well-tempered neutralino  
may be still consistent with the LUX constraints and the SM-like Higgs mass
provided that heavier Higgs particles, $H$ and $A$, are relatively light, 
with masses below about 400 GeV. Such region of the parameter space 
arises due to destructive interference between the $h$- and $H$-exchange 
contributions to the SI LSP-nucleon scattering amplitude. This small corner of
the parameter space is expected to be covered by the LHC searches for 
MSSM Higgses in the $\tau\tau$ decay channel with the data already 
collected in 2016 that are being analyzed.

We also discuss a well-tempered neutralino dominated by higgsino and show 
that the 2016 LUX results pushed the lower mass limit of such LSP to about 
950 GeV under the assumption of relatively light MSSM stops. In the 
presence of light MSSM Higgs particles this limit is relaxed to about 900 GeV.

This paper is organized as follows. In section \ref{sec:Omega} we review 
how a well-tempered neutralino may obtain the correct thermal relic
abundance. In section \ref{sec:LUXh}, we analyze current and future 
constraints on a well-tempered neutralino with a special emphasizes on 
the dependence on $\tan\beta$ and derive lower bounds on the MSSM stop 
masses as a function of the LSP mass. In section \ref{sec:LUXhH} we discuss 
the impact of not-decoupled heavy Higgs doublet on the LUX constraints 
and resulting relaxed lower bounds on the stop mass scale. We reserve 
section \ref{sec:concl} for concluding remarks on our results.

\section{Thermal relic abundance of well-tempered neutralino}
\label{sec:Omega}

We start with a review of the MSSM well-tempered neutralinos. 
In the present work we focus on well-tempered neutralinos which 
are such mixtures of bino and higgsinos which give the value of thermal 
relic abundance as inferred from experiments without substantial 
contribution from any resonances or co-annihilations other than 
those with the neutralinos and charginos. In such a case the thermal 
relic density is solely determined by the LSP mass and composition.

After decoupling of wino (assumed in this work) 
the neutralino mass sub-matrix describing the three lightest
states takes the form:
\begin{equation}
\label{M_chi}
 {M_{\chi^0}}=
\left(
\begin{array}{ccc}
  M_1 & -M_Z s_W \cos\beta & M_Z s_W \sin\beta \\[4pt]
  -M_Z s_W \cos\beta & 0 & -\mu  \\[4pt]
  M_Z s_W \sin\beta  & -\mu & 0  \\
\end{array}
\right) \,,
\end{equation}
where $s_W\equiv \sin\theta_W$. Trading $M_1$
for one of the eigenvalues, $m_{\chi_j}$, of the above neutralino mass
matrix we find the following (exact at the tree level) relations 
for the neutralino diagonalization matrix 
elements:
\begin{align}
\label{Nj3Nj1}
\frac{N_{j3}}{N_{j1}}
=
-
\frac{M_Z s_W}{\mu}
\,
\frac{(m_{\chi_j}/\mu)\sin\beta+\cos\beta}
{1-\left(m_{\chi_j}/\mu\right)^2}
\,,\\[4pt]
\label{Nj4Nj1}
\frac{N_{j4}}{N_{j1}}
=
-
\frac{M_Z s_W}{\mu}
\,
\frac{(m_{\chi_j}/\mu)\cos\beta+\sin\beta}
{1-\left(m_{\chi_j}/\mu\right)^2}
\,,
\end{align}
where $N_{j3}$, $N_{j4}$ and $N_{j1}$ denote, respectively, the two higgsino 
and the bino components of the $j$-th neutralino mass eigenstate
while $j=1,2,3$ and $|m_{\chi_1}|\le|m_{\chi_2}|\le|m_{\chi_3}|$.
The last two equations are enough to fully determine the composition 
of the three lighter neutralinos ($N_{j2}\approx0$ for decoupled wino)
and express it in terms of:  $M_Z/\mu$, $m_{\chi_j}/\mu$ and $\tan\beta$.
Later we will be interested mainly in the LSP corresponding to $j=1$, 
so to simplify the notation we will use $m_{\chi}\equiv m_{\chi_1}$. 
The physical (positive) LSP mass is given by $m_{\rm LSP}\equiv|m_{\chi}|$.

\begin{figure}
\center
\includegraphics[width=0.49\textwidth]{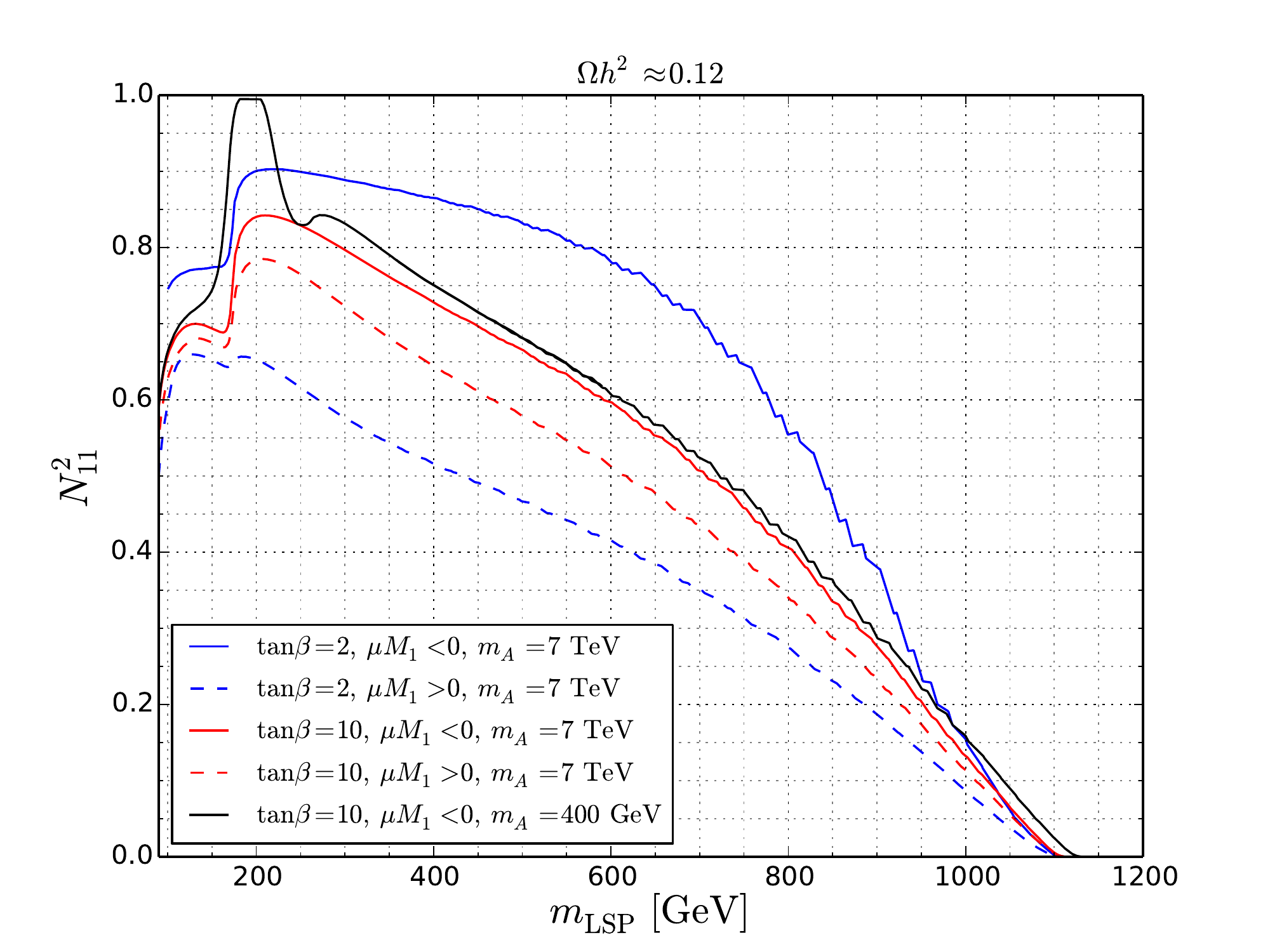}
\includegraphics[width=0.49\textwidth]{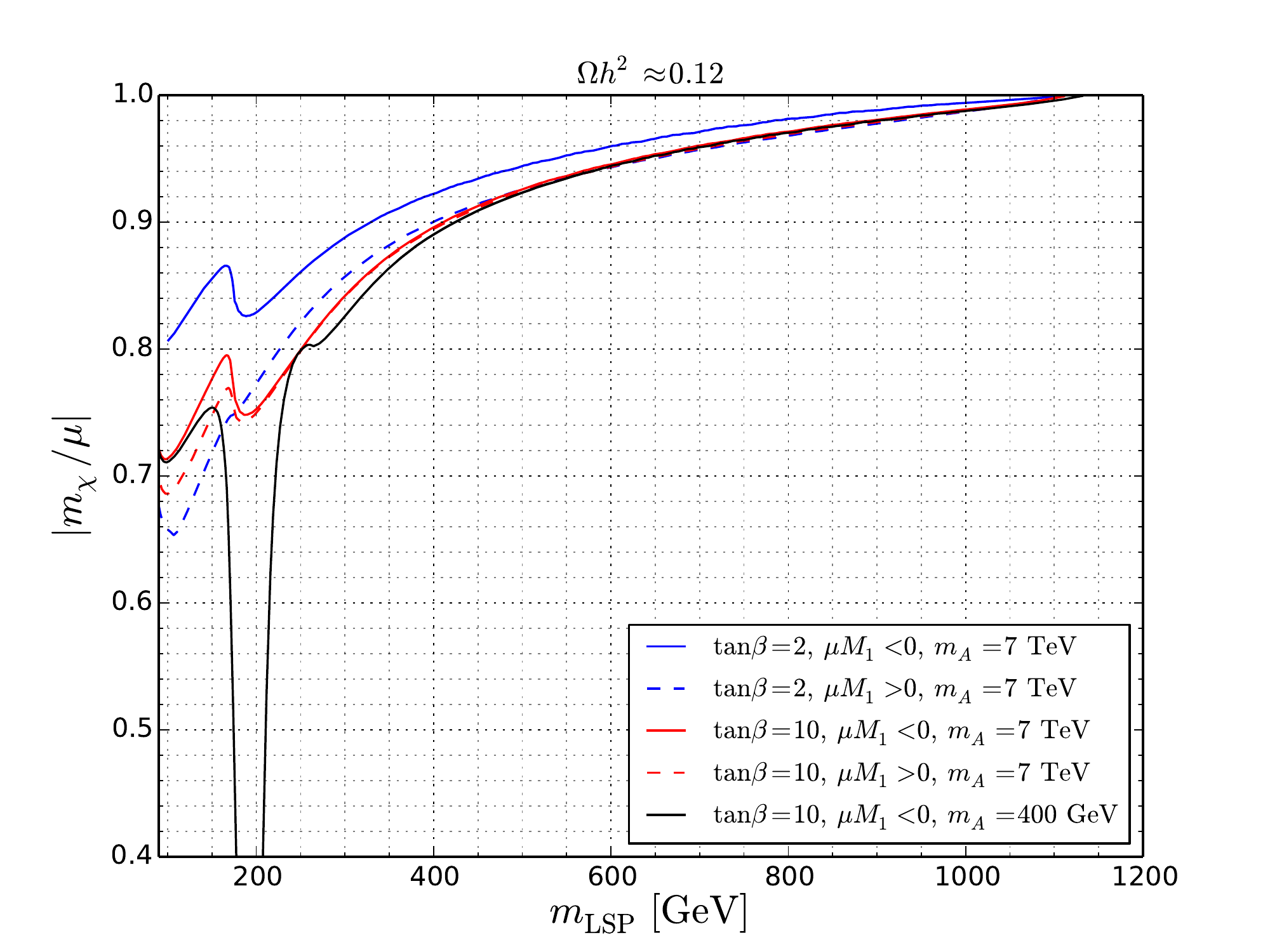}
\caption{Contours of $\Omega h^2\approx 0.12$ for several different 
values of $\tan\beta$, ${\rm sgn}(\mu M_1)$ and $m_A$ in the plane 
$m_{\rm LSP}$-$N_{11}^2$ (left panel) and $m_{\rm LSP}$-$|m_{\rm \chi}/\mu|$ 
(right panel). Note that in the right panel $\mu$ is evaluated at the 
$M_Z$ scale (in order to have $|m_\chi/\mu|\leq 1$) in contrary to the 
rest of the work (see right panel in Fig.~\ref{fig:blad_omega} and 
Table~\ref{tab:benchmarks}) where $\mu$ is evaluated at the SUSY scale.
}
\label{fig:mLSP_N11}
\end{figure}

Under the assumptions specified above $\Omega h^2\gg 0.12$ for the pure bino.  
On the other hand, the pure higgsino LSP annihilates very efficiently into 
$W^+W^-$ and $t\bar{t}$. Moreover, since the higgsino neutralino is almost  
degenerate with the chargino the co-annihilation is also very important. 
All these imply that the pure higgsino has too small relic abundance 
to explain all of the dark matter of the Universe if its mass is below 
about 1.1 TeV. Smaller masses of the LSP are possible only if a non-negligible 
bino component of the LSP is also present. The amount of required bino 
admixture depends on the LSP mass, as well as on $\tan\beta$ and 
${\rm sgn}(\mu M_1)$, as can be seen from the left panel of 
Fig.~\ref{fig:mLSP_N11}.
Throughout the paper, we calculate the MSSM spectrum using
{\tt SuSpect 2.41} \cite{Suspect} unless stated otherwise while we
always use {\tt MicrOMEGAs 4.3.1} \cite{micromegas} to calculate the
relic abundance and LSP scattering cross-section on nuclei.
 In our numerical calculations (with one exception mentioned later)
  we choose the gaugino mass parameters $M_2=M_3=7$~TeV.

For the following discussion it is useful to introduce distinction between 
two regions of the well-tempered neutralino parameter space:
\begin{itemize}
 \item $N_{11}^2\geq 0.5$  $\Leftrightarrow$ well-tempered bino-higgsino  
 \item $N_{11}^2 < 0.5$  $\Leftrightarrow$ well-tempered higgsino-bino  
\end{itemize}
The above two types of LSP are qualitatively different in several respects. 
As can be seen from Fig.~\ref{fig:mLSP_N11}, a well-tempered bino-higgsino
is generally lighter than a well-tempered higgsino-bino.
The transition occurs for $m_{\rm LSP}$ in the range of $700\div900$ 
($400\div600$) GeV 
for negative (positive) $\mu M_1$, as seen in Fig.~\ref{fig:mLSP_N11}.
The limiting case of a well-tempered higgsino-bino is the pure higgsino 
with mass of about 1.1 TeV while there is no corresponding limiting case for
a well-tempered bino-higgsino since the pure bino leads to overabundance. 
For a bino-higgsino the annihilation to $t\bar{t}$ is the dominant mechanism 
that washes out LSP in the Early Universe, while for a higgsino-bino it is 
the co-annihilation between the higgsino states.

For the discussion of direct detection constraints in the next section, 
it is also important to note that for a well-tempered higgsino-bino 
$|m_{\rm\chi}/\mu|\gtrsim 0.95\, (0.9)$ if $\mu M_1$ is negative (positive), 
as seen from the right panel of Fig.~\ref{fig:mLSP_N11}. For a well-tempered
bino-higgsino $|m_{\rm\chi}/\mu|$ can be lower but still above about 0.7 (0.65) 
for negative (positive) $\mu M_1$.

\section{Scattering of dark matter on nuclei}
\label{sec:LUXh}

The lightest neutralino in MSSM has spin-independent interactions with nuclei 
via exchange of squarks and Higgs scalars and spin-dependent (SD) ones via
exchange of $Z$ boson and squarks. The SI cross-section is given by
\begin{equation}
\sigma_{\rm SI}
=
\frac{4\mu^2_{\rm red}}{\pi}\,\frac{\left[Zf^{(p)}+(A-Z)f^{(n)}\right]^2}{A^2}
\,,
\end{equation}
where $\mu^2_{\rm red}$ is the reduced mass of the nucleus and the LSP. 
The effective couplings $f^{(N)}$ ($N=p,n$) are dominated by the t-channel 
exchange of the CP-even scalars
\cite{Jungman:1995df}:
\begin{equation}
\label{fN}
f^{(N)}
\approx
\sum_{i=1}^2
f^{(N)}_{h_i}
\equiv
\sum_{i=1}^2
\frac{\alpha_{h_i\chi\chi}\alpha_{h_iNN}}{2m_{h_i}^2}
\,,
\end{equation}
where $h_i=h,H$ are two CP-even scalar mass eigenstates.
The contributions from squarks have been neglected which is a good 
approximation due to the strong LHC constraints on their masses. In the limit of decoupled non-SM-like scalar, $H$, the SI scattering 
cross-section is entirely determined by the LSP mass and its coupling 
to the SM-like Higgs, $h$, given by the following function of the LSP 
composition:
\begin{equation}
\alpha_{h\chi\chi}
\approx
\sqrt{2}g_1 
\left[
N_{11}\left(N_{13}\sin\beta+N_{14}\cos\beta\right)
\right]
\,,
\label{alpha-hchichi}
\end{equation}
where we neglected the mixing between $h$ and $H$ which must be small 
in order to comply with the LHC Higgs measurements \cite{Higgscomb}. 
Using eqs.~\eqref{Nj3Nj1} and \eqref{Nj4Nj1} this coupling may be 
rewritten in the form
\begin{equation}
\label{alpha-hchichi_mumchi}
\alpha_{h\chi\chi}
\approx -\sqrt{2}g_1 N_{11}^2 \frac{M_Z s_W}{\mu} 
\,\frac{m_\chi/\mu + \sin(2\beta)}{1-\left(m_{\chi}/\mu\right)^2}
\,.
\end{equation}
Since $\sgn(m_{\chi})=\sgn(M_1)$ the above formula shows that the SI scattering 
cross-section is larger for positive $\mu M_1$ than in the opposite case
since there is no (partial) cancellation between the two terms in the numerator 
of eq.~\eqref{alpha-hchichi_mumchi}.
For negligible $h$-$H$ mixing the Higgs-nucleon coupling depends only
on $\tan\beta$:
\begin{equation}
\label{alpha_hNN}
\alpha_{hNN}
\approx
\frac{m_N F_d^{(N)}}{\sqrt{2}v\cos\beta}\,,
\end{equation}
where the form factors may be approximated by $F_d^{(p)}\approx0.132$, 
$F_d^{(n)}\approx0.140$ \cite{micromegas}.

Before we present our results one comment regarding our numerical procedure to compute the SI scattering cross-section is in order.
For small values of $\tan\beta$ {\tt SuSpect} is not able to compute the Higgs mass for very heavy stops required
to get the 125 GeV Higgs mass. Therefore, in our calculation of the SUSY spectrum (that is an input for {\tt MicrOMEGAs}) we set all squark masses to
6 TeV and rescale the SI scattering cross-section obtained from {\tt MicrOMEGAs} by a factor  $(m_h^{\rm M}/125\ {\rm GeV})^4$, where $(m_h^{\rm M}$
is the output of
the Higgs mass from {\tt MicrOMEGAs} (calculated by {\tt SuSpect}).

\begin{figure}[t]
\center
\includegraphics[width=0.49\textwidth]{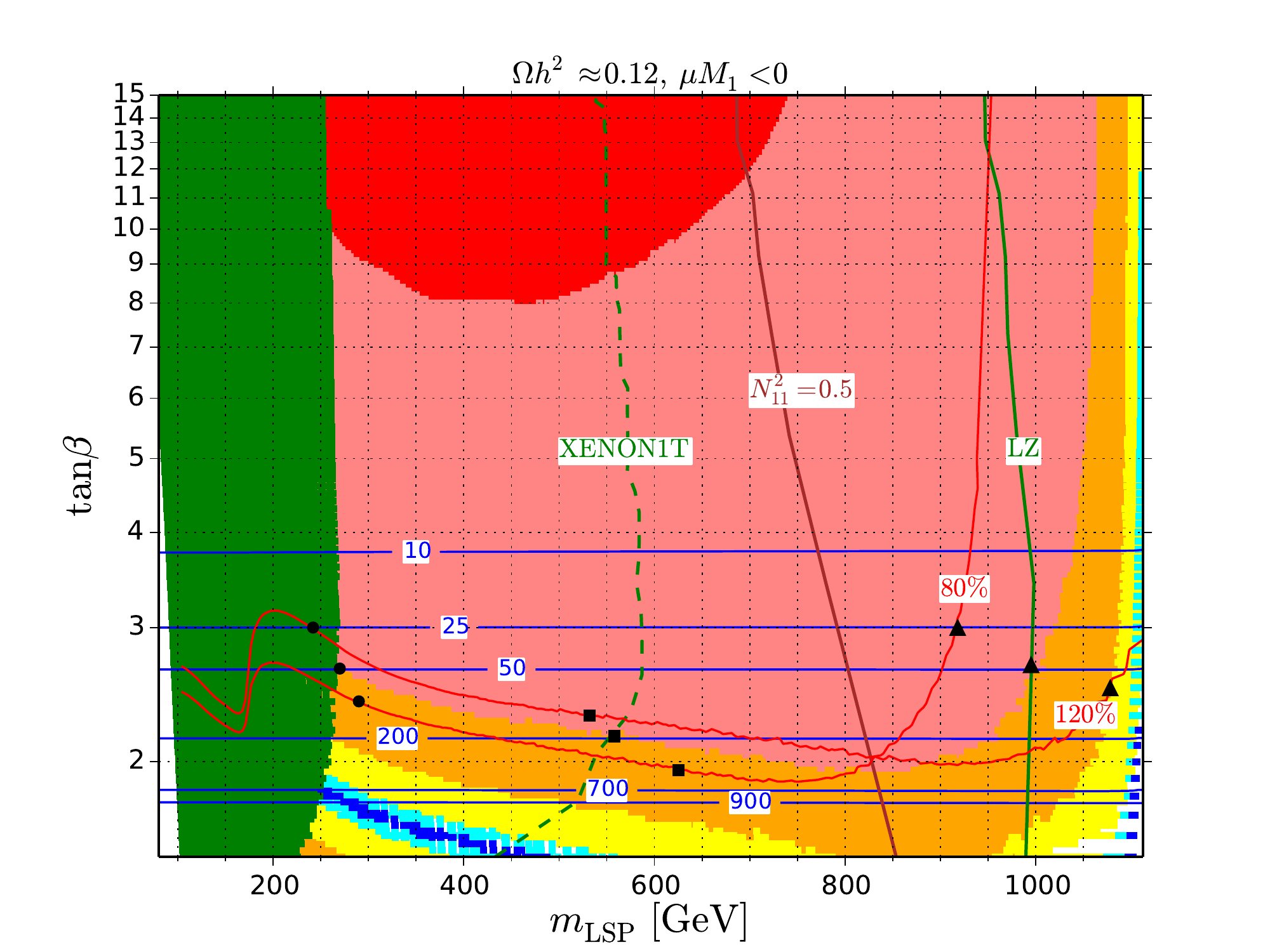}
\includegraphics[width=0.49\textwidth]{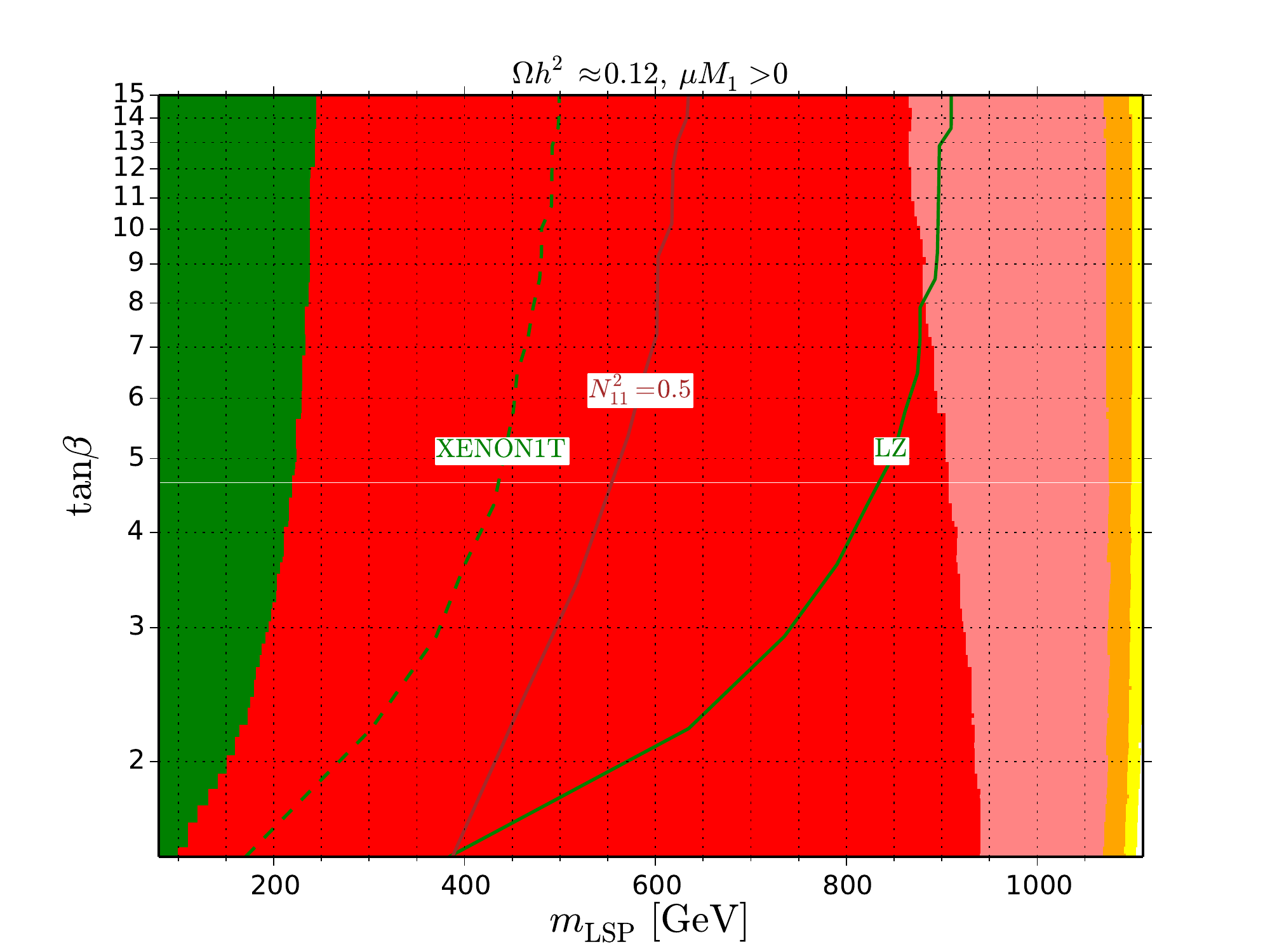}
\caption{Constraints on the well-tempered neutralino with $\Omega h^2\approx 0.12$ from current and future DD experiments in the plane $m_{\rm
LSP}$-$\tan\beta$. Light (dark) red region is excluded by the new (previous) LUX constraints on the SI scattering cross-section. Orange (yellow)
regions
are currently allowed but can be probed by future XENON1T (LZ) experiment. The light blue region is below the LZ sensitivity but above the
irreducible neutrino background (NB), whereas the dark blue region is the SI blind spot region with the cross-section below NB. Green 
region is excluded by the new LUX constraints on the SD scattering 
cross-section~\cite{LUX_SD}.  
The green lines denote future XENON1T and LZ sensitivity to SD scattering 
cross-section~\cite{SD_future}.
For instance the LZ experiment is expected to probe the entire parameter space
of the well-tempered bino-higgsino, 
which is to the left of the brown line with $N_{11}^2=0.5$. 
The blue lines correspond to minimal value of $M_{\rm SUSY}$ in TeV 
required to accommodate the measured Higgs mass of 125 GeV.
The red contours show the present constraints from the 2016 LUX data 
on $\sigma_{\rm SI}$ for $\Omega h^2$ changed by 20\%. The black symbols 
are related to the present constraints on and future sensitivity to 
SD cross-section (see Fig.~\ref{fig:blad_omega} for more details).
}
\label{fig:sigma_SI_tanbmLSP}
\end{figure}

As can be seen from the above equations, the SI LSP scattering cross-section on nuclei depends 
on three parameters: $\mu$, $m_\chi$ and $\tan\beta$. The requirement that 
such LSP accounts for the observed DM density fixes one of these relevant 
parameters. This allows us to plot constraints from direct detection 
experiments in a two-dimensional plane. From our perspective it is most 
interesting to choose the LSP mass and $\tan\beta$ as the two independent
variables. The current \cite{LUX2016} and previous LUX \cite{LUXold} constraints, as well as projections 
for XENON1T \cite{Xenon1T} and LZ \cite{LZ} experiments, for both signs of $\mu M_1$ are presented
in Fig.~\ref{fig:sigma_SI_tanbmLSP}. 
One can see that, as expected from 
eq.~\eqref{alpha-hchichi}, the SI scattering cross-section is larger for 
positive $\mu M_1$ than in the opposite case.  For positive $\mu M_1$ 
a well-tempered bino-higgsino was already excluded by the previous LUX 
results while the lower limit on the mass of a well-tempered higgsino-bino 
was improved by the new LUX results by more than 100 (200) GeV for small 
(large) values of $\tan\beta$. The present lower bound is almost 
$\tan\beta$-independent. Its precise position depends on the size of 
assumed uncertainties of the relic abundance calculation. For the 
values of $\Omega$ returned by {\tt MicrOMEGAs 4.3.1} this bound is about 
1070 GeV. It drops by about 100 GeV if 20\% theoretical uncertainty 
in $\Omega h^2$ is allowed.
Independently of these theoretical uncertainties, 
for $\mu M_1>0$ only an almost pure higgsino is allowed. Some region of 
higgsino-bino will never be probed by DD experiments 
since the SI scattering cross-section for a pure higgsino is only loop 
generated and is below the neutrino background due to some cancellations 
between the one- and two-loop diagrams 
\cite{Hisano:2011cs,Hill:2011be}.\footnote{
Some part of this region of the parameter space may be covered by 
future indirect detection experiments such as Cherenkov Telescope Array
\cite{Roszkowski:2014iqa}.}

The case of negative $\mu M_1$ is less constrained because 
of the existence of a blind spot in which the LSP-Higgs coupling 
vanishes in spite of non-vanishing higgsino-bino mixing \cite{bs_Hall}:
\begin{equation}
\label{bs:h}
 \frac{m_\chi}{\mu}= -\sin(2\beta) \,.
\end{equation}
However, $|m_\chi|$ can not be much smaller than $|\mu|$ if the
thermal relic abundance of the LSP is to be close to the value 
measured by Planck. Thus, the above blind spot occurs for very low 
values of $\tan\beta$: close to 1 for a well-tempered higgsino-bino 
and at most 2 for a well-tempered bino-higgsino.
The SI cross-section increases when we move in the parameter space 
away from a blind spot e.g.~by increasing $\tan\beta$. 
However, $\sigma_{\rm SI}$ may stay below a given DD experimental bound if 
we do not move too far. The allowed range of $\tan\beta$ values 
shrunk substantially after the latest LUX results. 
While the previous LUX results allowed a well-tempered bino-higgsino 
with $\tan\beta$ as large as about 10, the 2016 results set an upper 
bound of about 
2.7 after taking into account also the latest LUX constraints 
on $\sigma_{\rm SD}$ which exclude the LSP mass below about 260 GeV. Point P1 in Table~\ref{tab:benchmarks} is a representative benchmark for
well-tempered bino-higgsino with the SI and SD scattering cross-section just below the LUX upper limits.
\footnote{The constraints on $\sigma_{\rm SD}$ from PandaX \cite{Panda_SD_n} 
and IceCube \cite{IceCubeNEW} are slightly weaker than the latest LUX
constraints. }
The sensitivity of XENON1T experiment to the SI interactions may 
be high enough to move further this upper bound to about $2.2$.

Constraints on the SD scattering cross-section provided by
near future experiments will be a powerful complementary probe of a 
well-tempered bino-higgsino. XENON1T may improve the upper bound on $\tan\beta$
to about 1.7. 
when constraints on both SI and SD scattering cross-section 
are taken into account. The future LZ sensitivity may be enough to probe
the remaining part of the parameter space of well-tempered bino-higgsino.

\begin{figure}[t]
\center
\includegraphics[width=0.49\textwidth]{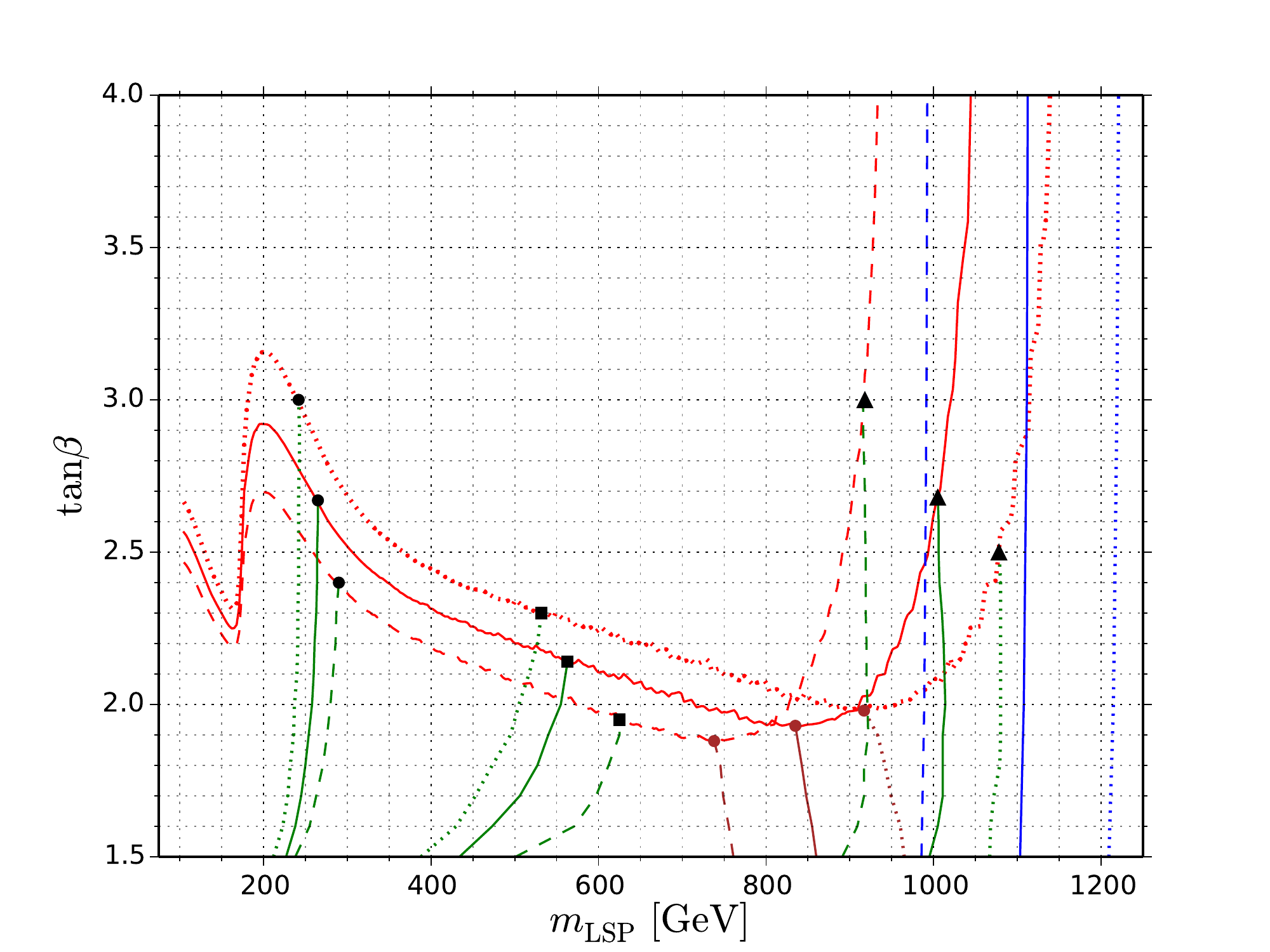}
\includegraphics[width=0.49\textwidth]{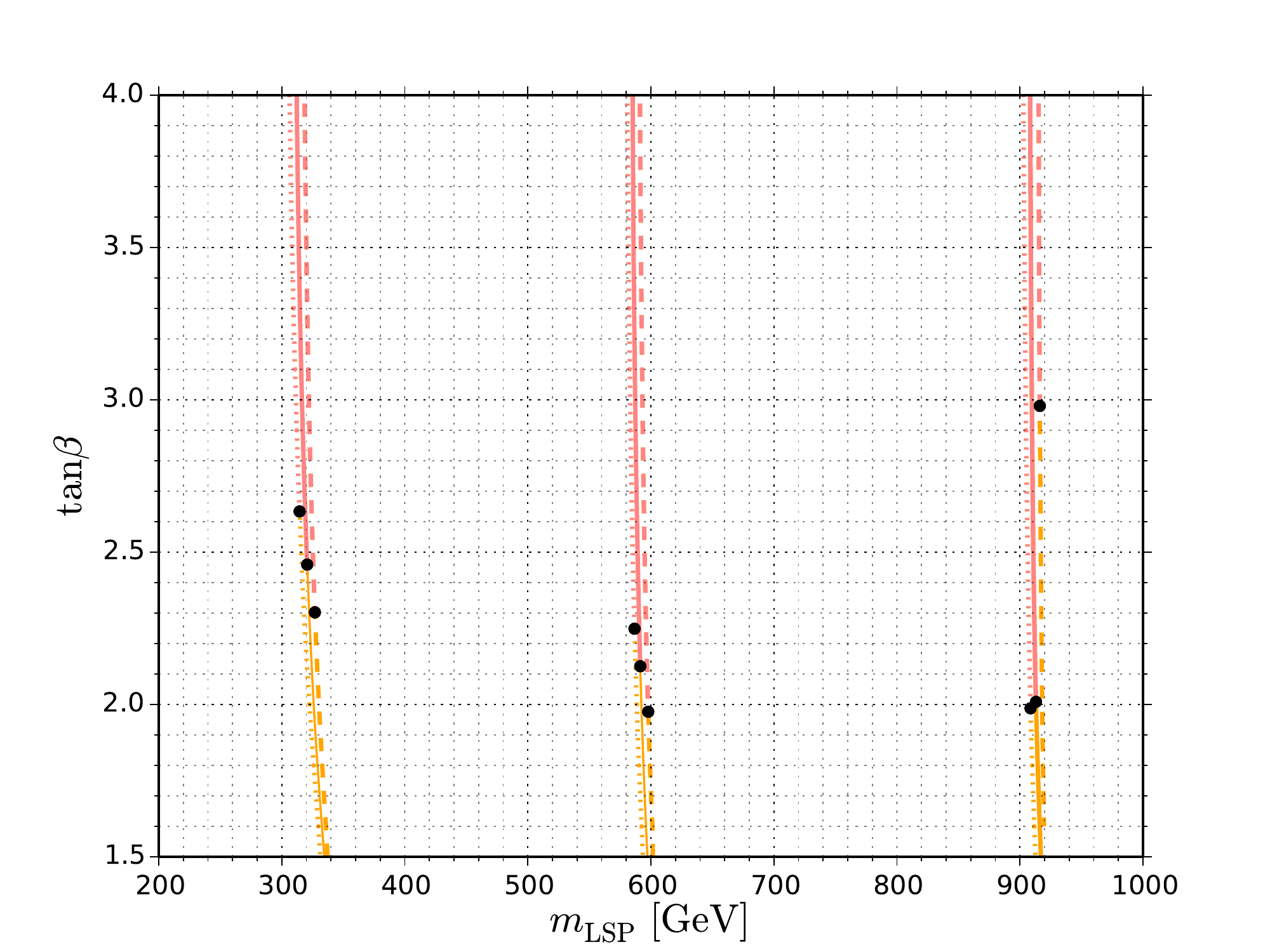}
\caption{Sensitivity of the constraints in the $m_{\rm LSP}$-$\tan\beta$ 
plane for $\mu M_1<0$ related to the uncertainty of $\Omega h^2$. 
All solid lines 
were obtained for $\Omega h^2\approx 0.12$ calculated with {\tt MicrOMEGAs 4.3.1}. 
All dashed (dotted) lines correspond to $\Omega h^2$ smaller (bigger) 
by 20\%.
Left panel: regions above the red contours are excluded by the 2016 LUX  
SI data. Green contours show sensitivity of present and future 
experiments to SD cross-section. Regions to the left of these contours 
are excluded or will be probed by (from left to right) LUX, XENON1T and 
LZ experiments. Black symbols denote crossing of the corresponding red and
green lines (and are shown also in Fig.~\ref{fig:sigma_SI_tanbmLSP}).
Brown lines correspond to the bino-higgsino border with $N_{11}^2=0.5$.
Blue lines show maximal (pure-higgsino) LSP mass for which assumed 
values of $\Omega h^2$ may be achieved.
Right panel: lines of fixed values of the $\mu$ parameter equal 
(from left to right) 350, 600 and 900 GeV. Parts above (below) the black 
dots are excluded (allowed) by the LUX 2016 data on $\sigma_{\rm SI}$.
 }
\label{fig:blad_omega}
\end{figure}

All the above discussed (and color-coded in Fig.~\ref{fig:sigma_SI_tanbmLSP})
constraints in the $m_{\rm LSP}$-$\tan\beta$ plane were obtained for such 
points in the MSSM parameter space for which $\Omega h^2$  
calculated with {\tt MicrOMEGAs 4.3.1} equals 0.12. The experimental 
error on $\Omega h^2$ is quite small but there are some theoretical 
uncertainties in the corresponding calculations. In order to estimate 
the sensitivity of our results to such uncertainties we repeated 
our analysis allowing for $\pm20$\% errors on $\Omega h^2$. In Fig.~\ref{fig:blad_omega} we present the impact of the uncertainty in $\Omega h^2$ 
on current and future experimental constraints from DD.  
We also showed the impact of uncertainty in $\Omega h^2$ on the most 
relevant constraints in Fig.~\ref{fig:sigma_SI_tanbmLSP}. 
These uncertainties do not affect our general conclusions but they 
somewhat change numerical values for the bounds quoted above. 
In particular, the upper bound on $\tan\beta$ can be relaxed to about 3. 

\begin{table}
\centering
\begin{tabular}{|l|ccc|}
\Xhline{2\arrayrulewidth}
& {\rm P1} & {\rm P2} & {\rm P3}   \\
\hline\hline
$\mu$ [GeV] & -310 & -1032 & -346   \\
$M_1$ [GeV] & 280 & 1137 & 306 \\
$M_2$, $M_3$ [GeV] & 7000 & 7000 & 2000 \\
$\tan \beta$ & 2.6 & 4 & 6.9 \\
$m_A$ [GeV] & 6000 & 6000 & 350 \\
\hline
$m_{h}$ [GeV] & 125 & 125 & 125.5 \\
$m_{\rm LSP}$ [GeV] & 275 & 1050 & 300    \\
$m_{\tilde{\chi}^\pm_1}$ [GeV] & 322 & 1056 & 353 \\
$M_{\rm SUSY}$ [TeV] & 54 & 8.5 & 1.5   \\
\hline
$\Omega h^2$  & 0.12 & 0.12 &  0.12 \\
$\sigma_{\rm SI}$ [cm$^2$] & $3.9\times 10^{-46}$ & $1.4\times 10^{-45}$ & $4.3\times 10^{-46}$ \\
$\sigma_{\rm SI}^{\rm LUX}$ [cm$^2$] & $4.0\times 10^{-46}$ & $1.6\times 10^{-45}$ & $4.4\times 10^{-46}$  \\
$\sigma_{\rm SD}$ [cm$^2$] & $6.6\times 10^{-41}$ & $3.1\times 10^{-42}$ & $6.4\times 10^{-41}$ \\
$\sigma_{\rm SD}^{\rm LUX}$ [cm$^2$] & $7.1\times 10^{-41}$ & $2.5\times 10^{-40}$ & $7.7\times 10^{-41}$ \\
\Xhline{2\arrayrulewidth}
\end{tabular}
\caption{List of benchmark points with $\Omega h^2\approx 0.12$ that have the SI scattering cross-section just below the current LUX bound
$\sigma_{\rm SI}^{\rm LUX}$. For points P1 and P2, the value of $M_{\rm SUSY}$ that gives $m_h=125$ GeV was computed with {\tt SUSYHD}, while for P3
the whole SUSY spectrum is computed with {\tt SuSpect}.
} 
\label{tab:benchmarks}
\end{table}

In the right panel of Fig.~\ref{fig:blad_omega} we show how this upper 
bound depend on $\Omega h^2$ for fixed values of $\mu$ parameter. 
Typically it is relaxed (strengthened) by less than about 0.2 when 
$\Omega h^2$ is 20\% bigger (smaller). 
The only exception is for large values of $\mu$ which is related 
to the fact that the relic density of (almost) pure higgsino depends 
quite strongly on its mass so also on $\mu$.
It is also clear from Fig.~\ref{fig:blad_omega} that relation between 
value of $\mu$ and the LSP mass very weakly depends on  
uncertainties in $\Omega h^2$.

A strong upper bound on $\tan\beta$ for a well-tempered bino-higgsino 
implies a strong lower bound on the stop masses since the radiative 
corrections to the Higgs mass must be huge to account for the observed 
value of 125 GeV \cite{Higgsmassexp}. In order to compute such lower 
bounds in the region of small $\tan\beta$ we used a program {\tt SUSYHD 1.0.2} \cite{SUSYHD} which
calculates the Higgs mass in the MSSM using the Effective Field Theory 
framework which gives much more reliable results than the standard SUSY 
spectrum calculators in the region of multi-TeV stop
masses.\footnote{
  Typical estimate of theoretical uncertainty by {\tt SUSYHD} is about 1 GeV.
  In our calculation of a lower bound on the stop masses we use the central
  value of the Higgs mass returned by {\tt SUSYHD}.
}
In Fig.~\ref{fig:sigma_SI_tanbmLSP}, contours of 
lower bounds on $M_{\rm SUSY}\equiv\sqrt{m_{Q_3} m_{U_3}}\approx \sqrt{m_{t_1} m_{t_2}}$ in the plane of 
$m_{\rm LSP}$-$\tan\beta$ are presented. We see that the 2016 LUX results 
impose a lower bound on the stop masses of about 50 TeV (25 TeV if uncertainty on $\Omega h^2$ is taken into account)  
for the LSP mass of about 260 GeV and quickly gets stronger when going 
away from this
mass.\footnote{
  The exact value of this bound may vary by up to 30\% depending on
  the choice of $M_2$ and $M_3$ which do not influence dark matter
  phenomenology as long as wino is much heavier than higgsinos.
} 
Future XENON1T may improve the lower bound on the  
stop masses to about  900 TeV (for the LSP mass of about 500 GeV). 
As mentioned above, by the time the neutrino background is reached 
the well-tempered bino-higgsino should be already excluded or discovered 
by its SD scattering with nuclei at the LZ experiment.
In calculating these bounds we used the stop mixing value giving 
the maximal correction to the Higgs mass 
i.e.~$X_t\equiv A_t-\mu/\tan\beta=\sqrt{6} M_{\rm SUSY}$. 
Therefore, for generic stop mixing
the lower bound on the stop masses is much stronger.

Such a strong lower bound may be an indication for split SUSY \cite{splitSUSY1,splitSUSY2}. 
Interestingly the upper bound on $\tan\beta$ gives also a strong upper bound 
on the smuon mass if MSSM is responsible for the explanation of the 
long-standing muon $g-2$ anomaly \cite{Bennett:2006fi,Moroi:1995yh,Martin:2001st,g2}. This upper bound would force 
smuon to be the LSP which means that a well-tempered bino-higgsino is not 
consistent with the MSSM solution to the muon $g-2$ anomaly. This conclusion 
may change if the blind spot condition is modified in such a way that it 
may be fulfilled for $|m_\chi/\mu| \sim 1$ not only for small $\tan\beta$. 
We discuss this possibility in the next section.

The lower limit on the well-tempered higgsino-bino mass for values 
of $\tan\beta$ that allows for light MSSM stops and is consistent with 
2016 LUX results is similar as for the $\mu M_1>0$ case and is pushed 
to the region of almost pure higgsino. Since for the pure higgsino LSP the correct relic abundance is obtained for $m_{\rm LSP}\approx1.1$ TeV with
the uncertainty of about 100 GeV there is uncertainty of similar size in the lower bound on the well-tempered higgsino-bino mass. For example, for
$\tan\beta=4$, corresponding to a lower bound on stops masses of about 10 TeV, the lower bound on the mass of the LSP is about $1050\pm100$ GeV, as
seen from Fig.~\ref{fig:blad_omega}, see also point P2 in Table~\ref{tab:benchmarks}. There is also an upper bound 
slightly above $1100\pm100$ GeV due to the fact that even pure 
higgsinos have too large relic abundance if they are heavier.

If one insists on a natural realization of SUSY with rather light stops 
then a well-tempered bino-higgsino with decoupled MSSM-like Higgs boson
requires such extension of the MSSM which leads to additional contributions
to the Higgs mass for 
small $\tan\beta$. A prime example of such a model is the Next-To-Minimal 
Supersymmetric Standard Model (NMSSM) \cite{reviewEllwanger}. 
In NMSSM, there is much more flexibility to accommodate a well-tempered 
neutralino consistently with the current data because the light singlet
scalar which mixes with the Higgs may substantially modify the blind spot
condition \cite{BS_NMSSM} and the LSP can be also 
a mixture of singlino and higgsino \cite{MbMoPs_thermal}. However, as we 
show in the next section there is still some small region of the MSSM parameter 
space in which a well-tempered neutralino can be consistent with relatively 
light stops.

\section{Well-tempered neutralino with light $H$}
\label{sec:LUXhH}

The question that we would like to address in this section is whether 
a well-tempered bino-higgsino or higgsino-bino with a significant 
bino component may still be consistent with all constraints 
in the framework of MSSM with not very heavy stops. 
This may happen if the blind spot condition is modified in such a way 
that it has solutions with $|m_{\chi}|\approx |\mu|$ for not too small 
$\tan\beta$. It was pointed out in Ref.~\cite{bs_Wagner} that this may be 
possible provided that the heavier MSSM Higgs boson, $H$, is relatively light and its effective coupling to nucleons is large enough. 
The $H$ coupling to LSPs is approximately given by  
\begin{equation}
\alpha_{H\chi\chi}
\approx
\sqrt{2}g_1 
\left[
N_{11}\left(N_{14}\sin\beta-N_{13}\cos\beta\right)
\right]
=
\sqrt{2}g_1 N_{11}^2 \frac{M_Z s_W}{\mu} 
\,\frac{\cos(2\beta)}{1-\left(m_{\chi}/\mu\right)^2}
\,.
\label{alpha-Hchichi}
\end{equation}
The $H$ coupling to down quarks is enhanced by $\tan\beta$ which may compensate the mass-suppression of the effective $H$ coupling to nucleons.
The SI cross-section is then approximately proportional to
\begin{equation}
 \sigma_{\rm SI}\sim\left[ \frac{2}{m_h^2}\left( m_\chi + \mu \sin(2\beta) \right) + \mu \cos(2\beta) \left(-\tan\beta + 1/\tan\beta 
\right)\frac{1}{m_H^2} \right]^2 \,.
\end{equation}
Vanishing of the above cross-section leads at large $\tan\beta$ to the following blind spot condition:
\begin{equation}
\label{bs:hH}
 \frac{m_\chi}{\mu}\approx-\sin(2\beta)-\frac{m_h^2}{m_H^2}
\,\frac{\tan\beta}{2} \,.
\end{equation}
We see that also in this case the blind spot condition may be satisfied 
only for $\mu M_1<0$. A well-tempered neutralino implies $|m_\chi|\sim|\mu|$ 
so the r.h.s.~of the above equation has to be close to 1 in order for the 
SI cross-section to be strongly suppressed. The first term of the r.h.s.\ 
of \eqref{bs:hH} originates from the $h$-exchange amplitude and is 
suppressed if $\tan\beta$ is not small. The second term originates from 
the interference between the $h$ and $H$ contributions to the scattering 
amplitude. The $H$ contribution is enhanced by $\tan\beta$ so 
it may compete with the contribution from the lighter 
Higgs.\footnote{
Another contribution to the SI scattering amplitude which is 
enhanced by $\tan\beta$ may originate from light sbottoms \cite{bs_Crivellin}.
However, this contribution may compete with the Higgs contribution only 
for $|\mu|\gg|M_1|$ and/or very small mass splitting between bino and sbottom
that would lead to strong sbottom co-annihilations. Therefore, for 
a well-tempered neutralino this contribution can be safely neglected. 
}
It turns out, however, that the $H$ contribution is always smaller than the $h$ contribution due to the LHC constraints on the Higgs sector. 
The main constraint on this scenario comes from direct $H/A$ 
searches in the $\tau\tau$ decay channel that set a lower bound on 
$m_A\approx m_H$ as a function of $\tan\beta$ \cite{HAtautau_ATLAS,HAtautau_CMS}. 
The limits for $H/A$ masses are stronger for larger $\tan\beta$ and the 
second term on r.h.s.~of eq.~\eqref{bs:hH}, with constraints for $m_H$ 
saturated, decreases with $\tan\beta$. This is because for a given $m_A$ 
the production cross-section, dominated at large $\tan\beta$ by the bottom 
fusion mode, grows as $\tan^2\beta$. In the most relevant part of parameter space the CMS constraints are stronger than the ATLAS one mainly due to
some small deficit (excess) in the CMS (ATLAS) data for $m_A$ around 400 GeV.  
For $\tan\beta\lesssim7$ the production 
cross-section is small enough that the LHC searches in the $\tau\tau$ channel 
still allow relatively light $H$. However, it cannot be arbitrary light  
because small values of $m_A$ generically lead to an enhanced $hb\bar{b}$ 
coupling which results in suppressed Higgs signals in the gauge boson final 
states which are relatively well measured and found to be close to the SM predictions. Indeed, the ATLAS Higgs measurements set a lower bound on
$m_A$ 
of 375 GeV for moderate and large $\tan\beta$
\cite{ATLAS_MAlimit}. This limit, however, should be interpreted with care.  
The expected ATLAS lower limit on $m_A$ is about 315 GeV. The much stronger 
limit observed by ATLAS comes mainly from the fact that central values for 
$\gamma\gamma$ and $ZZ^*$ rates at Run I are above the SM prediction. 
The preliminary results of Run II data exhibit suppression of the 
$\gamma\gamma$ rate \cite{ATLAS_hgammagamma_ICHEP} so somewhat smaller 
values of $m_A$ than 375 GeV should not be considered as
definitely excluded. For example, during the recent Moriond conference CMS presented a result of $1.05\pm0.17$ for the $ZZ^*$ Higgs signal rate which
is experimentally the cleanest channel \cite{CMS_ZZ_Moriond}. This is the most accurate single measurement of the Higgs signal rate (even more precise
than the corresponding result of the ATLAS and CMS combination of the 8 TeV data) and we found that it sets a lower bound on $m_A$ at 95~\% CL of
about 320 GeV for moderate and large $\tan\beta$. 
Thus, there is complementarity between the Higgs properties measurements 
and the $\tau\tau$ searches at the LHC which lead to the upper bound on 
the r.h.s.\ of eq.~\eqref{bs:hH} of about 0.7 (0.8) if the ATLAS (conservative) limit of $m_A>375$ (320) GeV is used.  This is too small 
to put a well-tempered neutralino at a blind spot for values of 
$\tan\beta$ that allow light stops. On the other hand, it can be large 
enough to avoid the current LUX constraints.

\begin{figure}
\center
\includegraphics[width=0.49\textwidth]{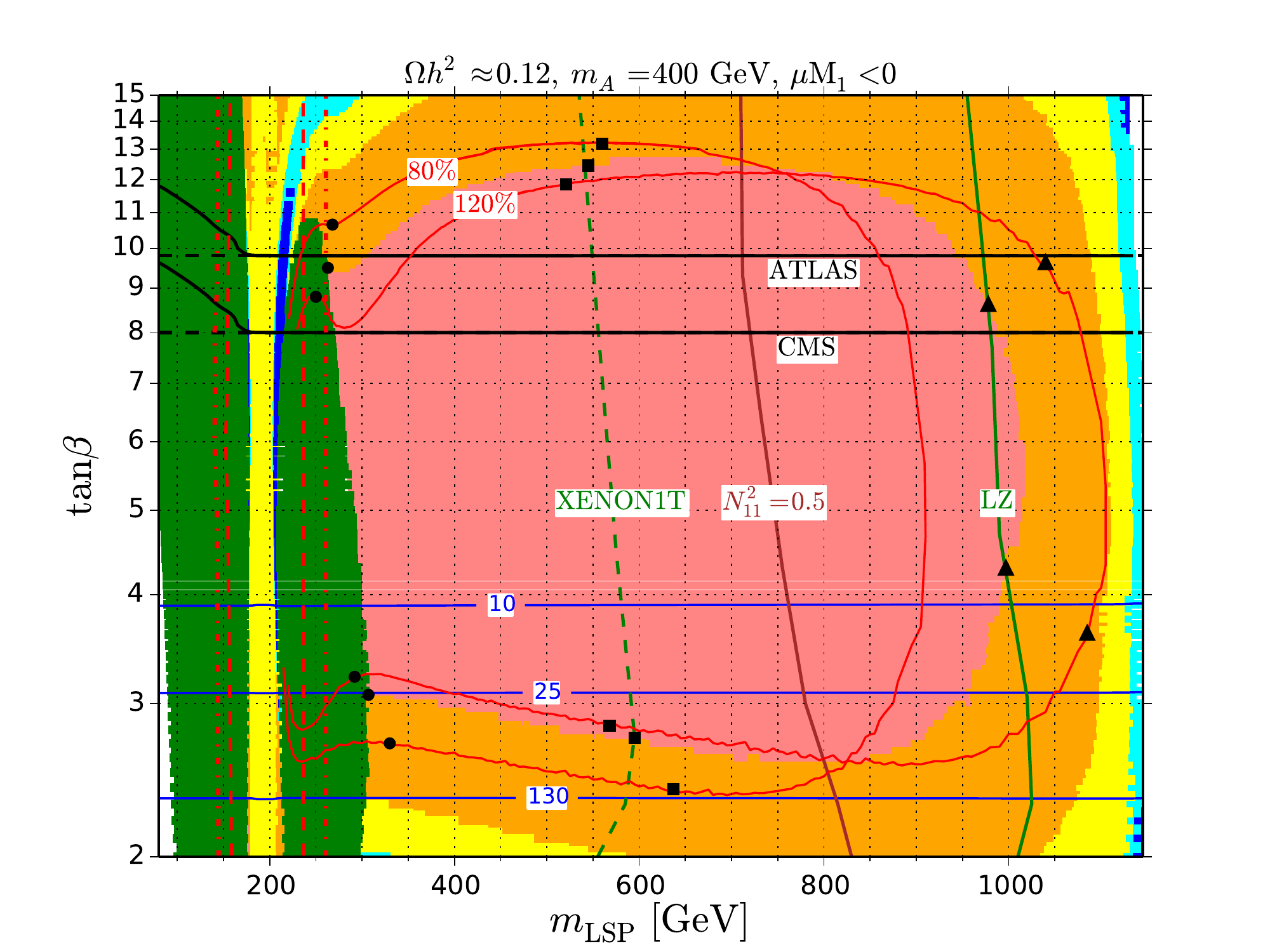}
\includegraphics[width=0.49\textwidth]{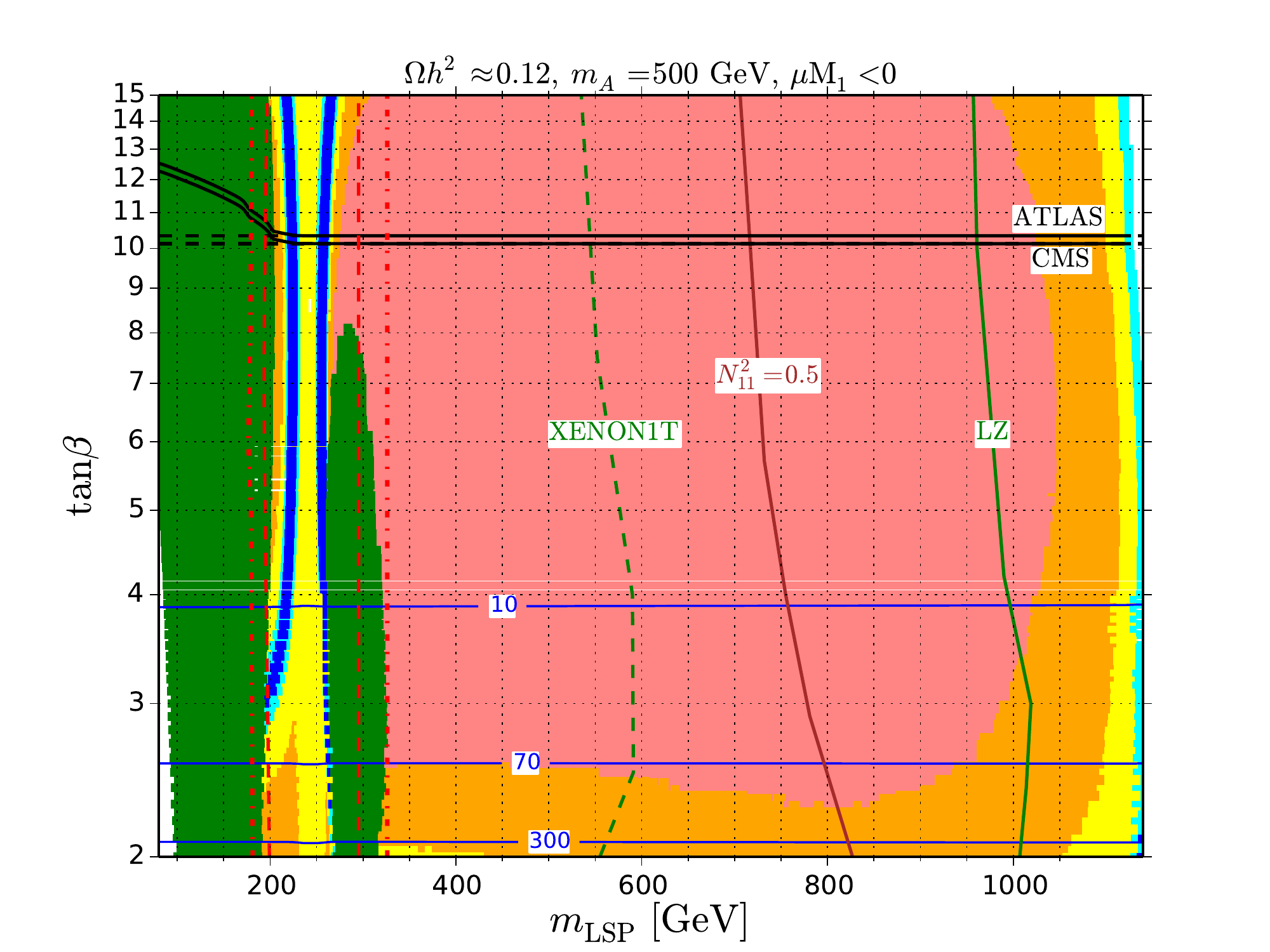}
\caption{Constraints on the well-tempered neutralino with 
$\Omega h^2\approx 0.12$ from current and future DD experiments in 
the plane $m_{\rm LSP}$-$\tan\beta$ for $m_A=400$ and 500~GeV. 
The color code is the same as in Fig.~\ref{fig:sigma_SI_tanbmLSP}. 
The black lines show the upper bounds on $\tan\beta$ obtained by 
ATLAS and CMS (see labels) searches for $H/A$ in the $\tau\tau$ decay 
channel (without taking into account $H/A$ decays to charginos and neutralinos).
  The regions between the red lines   (dashed or dashed-dotted)
  correspond to such parts of the parameter space where effects
  of resonant annihilation may be important (see Appendix for details). 
Meaning of all other contours and black symbols is as in 
Fig.~\ref{fig:sigma_SI_tanbmLSP}.
}
\label{fig:fixedMA}
\end{figure}

In Fig.~\ref{fig:fixedMA} we present the current and future constraints 
in the plane of $m_{\rm LSP}$-$\tan\beta$ for $m_A$ equal 400 and 500 GeV.
It is interesting how much the situation is changed by the LUX 2016 
bounds. No part of the parameter space shown in Fig.~\ref{fig:fixedMA} 
is inconsistent with the previous LUX results. On the other hand, 
the new results, together with the bounds from the $\tau\tau$ channel, 
exclude most of the shown parameter space for a bino-higgsino LSP 
and big part for a higgsino-bino LSP.

In the well-tempered bino-higgsino case $m_A$ as light as 400 GeV 
is still not sufficient to satisfy the new LUX constraints for moderate 
or large $\tan\beta$ without violating the CMS $\tau\tau$ constraints 
even after taking into account the uncertainty in $\Omega h^2$. 
Note also that decays of $H/A$ to the neutralinos and charginos 
can not relax the $\tau\tau$ constraints because such decays are 
relevant only for LSP masses below 200 GeV which are excluded by SD LUX 
constraints. For $m_A=400$ GeV the upper bound on small $\tan\beta$ is 
relaxed to slightly above 3 but this still requires very heavy stops.
For the central value of $\Omega h^2$ the lower bound on the stop masses 
are relaxed to about 25 GeV from 50 GeV in the case with decoupled $H/A$. 
With uncertainties in $\Omega h^2$ taken into account such relaxation 
is even weaker. 

Light $H$ and $A$ can also somewhat relax a lower bound on a well-tempered 
higgsino-bino LSP. Comparing Figs.~\ref{fig:sigma_SI_tanbmLSP} 
and \ref{fig:fixedMA} we see that the lower bound on $m_{\rm LSP}$ 
(for stop masses below 10 TeV) decreases from about 1050 GeV 
in the case of decoupled MSSM-like Higgses to about 1 TeV
for $m_A=400$ GeV. These numbers vary by about 100 GeV if $\Omega h^2$ is varied within 20~\% percent of the central value.

It is important to note that the presence of light $H$ and $A$ leads 
to resonant annihilation for $m_{\rm LSP}$ in the vicinity of $m_A/2$. 
In such a case a much smaller higgsino component, or equivalently much 
smaller $|m_\chi / \mu|$, is required to obtain $\Omega h^2\approx0.12$, 
as can be seen from Fig.~\ref{fig:mLSP_N11}. In consequence, close to the 
resonance the SI cross-section is much smaller and can be even below the 
neutrino background.
Such regions of very small SI cross-sections for $m_{\rm LSP}\sim m_A/2$
are clearly visible in both panels of Fig.~\ref{fig:fixedMA}.
However, they clearly violate the definition of a 
well-tempered neutralino. 
The criterion we use to separate the resonant
annihilation from a well-tempered neutralino is discussed 
in Appendix \ref{App:kryterium}.

\begin{figure}
\center
\includegraphics[width=0.49\textwidth]{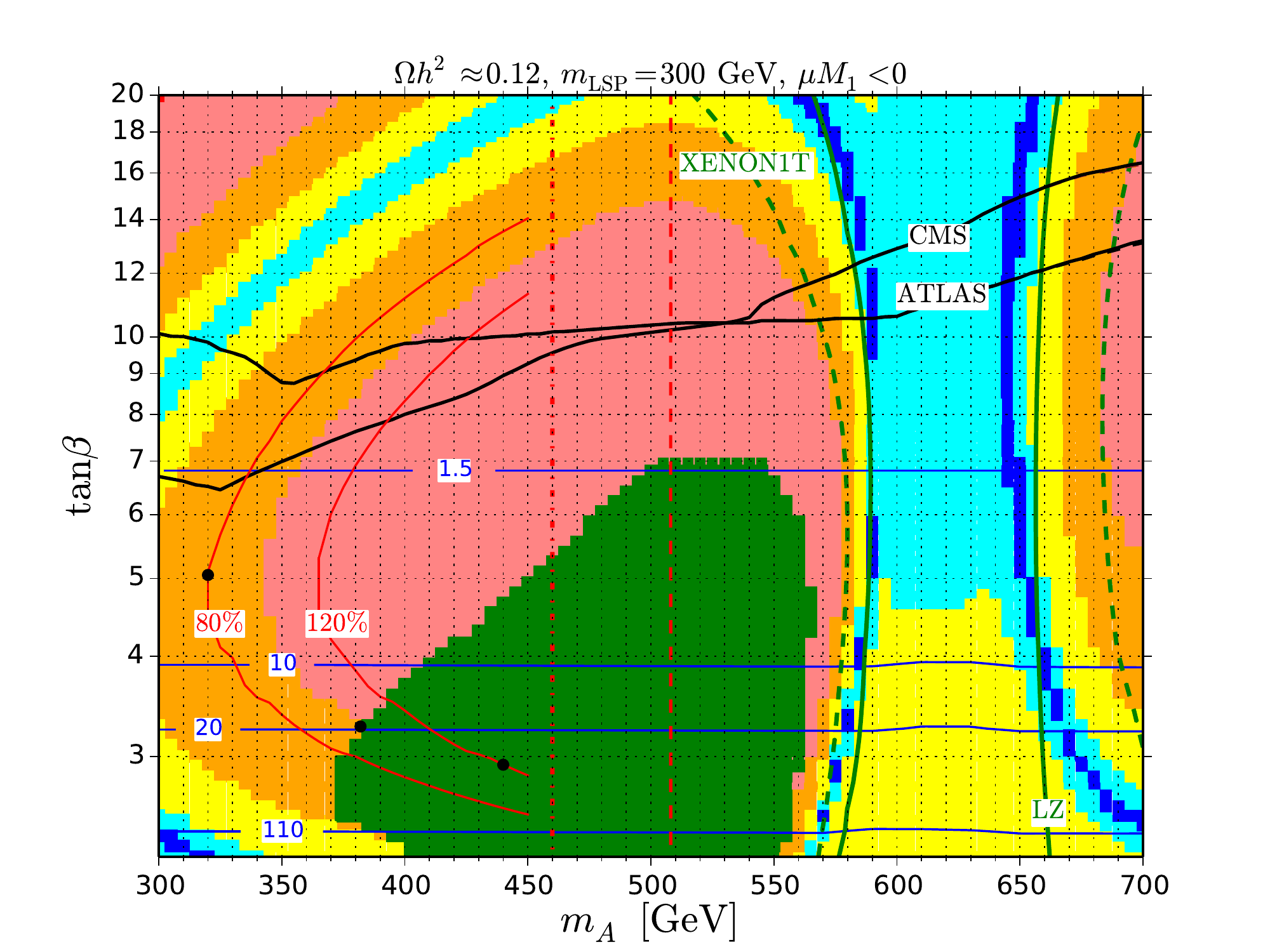}
\caption{Constraints on the well-tempered neutralino with 
$\Omega h^2\approx 0.12$ from current and future DD experiments 
in the plane $m_A$-$\tan\beta$
  for LSP mass of 300 GeV. The color code 
and meaning of all contours and symbols is the same as in 
Fig.~\ref{fig:fixedMA}.
  }
\label{fig:MAtanbeta}
\end{figure}
In order to obtain viable well-tempered bino-higgsino with relatively light stops one needs to consider $m_A$ below 400 GeV.
In Fig.~\ref{fig:MAtanbeta} we present current and future constraints in
the plane of $m_A$-$\tan\beta$ for the LSP mass of 300 GeV for which the new SI LUX constraints are the weakest. The constraints from the previous LUX
results were irrelevant for the presented parameter space.
The bounds become
much stronger when the 2016 LUX results are taken into account. 
For example, taking the ATLAS limit on $m_A>375$ GeV at face value 
and using the result for $\Omega h^2$ without theoretical uncertainty a
well-tempered bino-higgsino is excluded for moderate and large $\tan\beta$. 
If the uncertainty in $\Omega h^2$ is taken into account values of $m_A$ up
to about 390 GeV for $\tan\beta\approx8$ satisfy the latest LUX constraints 
on the SI cross-section with the stop masses not much bigger than 1 TeV.  
Point P3 in Table~\ref{tab:benchmarks} is a representative benchmark with relatively light stops that satisfies the $\tau\tau$ constraints, the
conservative limit on $m_A$ from the Higgs coupling measurements and has the SI scattering cross-section just below the LUX upper limit.
\footnote{We
  computed the Higgs mass at large (small) $\tan\beta$ with {\tt Suspect}
  ({\tt SUSYHD}).
   The line for $\tan\beta>5$ in Fig.~\ref{fig:MAtanbeta} was obtained for $M_2=M_3=2$~TeV
  (and not 7 TeV as in all other cases) because for large values of
  $\tan\beta$ lighter gauginos lead to slightly heavier Higgs without
  influencing strongly the LSP properties.}
 The whole allowed region is within reach of XENON1T. 
The LHC searches in the $\tau\tau$ channel may cover the remaining parts of
the parameter space even sooner than XENON1T experiment 
- already with the data collected in 2016 which is about 40 fb$^{-1}$. 
It is  noteworthy that  for $m_A$ below about 340 GeV (or 360 GeV if the uncertainty on $\Omega h^2$  is taken into account)  the SI LUX
constraints are satisfied for all $\tan\beta$ consistent
with the $\tau\tau$ constraints.
We should also note that the new SD LUX constraints rule out some range of $m_A$ for small and moderate $\tan\beta$, mostly in the transition region
between well-tempered neutralino and resonant annihilation. This is partly because in that region there is destructive interference in the LSP
annihilation into $t\bar{t}$ between the $Z$-mediated and $A$-mediated contributions so bigger higgsino component is needed to obtain correct relic
density.

\section{Conclusions}
\label{sec:concl}

We have systematically assessed the impact of 2016 LUX results on the MSSM
well-tempered neutralino that is a mixture of bino and higgsino which 
gives the correct thermal relic abundance without help of resonant annihilation
or co-annihilation with sfermions. We found that in the limit of
decoupled non-SM-like scalars the new LUX results set a very strong upper
bound on $\tan\beta$, hence also a strong lower bound on the stop
mass scale, if LSP mass is below about 1 TeV. 
For a well-tempered bino-higgsino $\tan\beta$ must be below about 3
which implies a lower bound on the stop mass scale of about 25 TeV when the
correction from stop mixing to the Higgs mass is maximized 
and when theoretical uncertainties in calculating DM relic abundance are taken 
into account.
Well-tempered higgsino-bino can still be compatible with light MSSM stops
but only for almost pure higgsino in a small range of masses between about
1050 and 1100 GeV. Uncertainties of $\Omega h^2$ may move this range
up or down by up to about 100 GeV.

The only way to save the well-tempered bino-higgsino with light MSSM stops
is to consider the MSSM-like Higgs doublet to be relatively light.
In such a case the contribution to the SI LSP-nucleon
interaction from the $H$ exchange may partially cancel the SM-like Higgs
contribution.
We have shown that the latest LUX constraints on the well-tempered bino-higgsino
can be satisfied with $\tan\beta$ up to about 8 in a small part of the
parameter space that will be entirely probed by XENON1T.
Accommodating the LUX constraints requires $m_A$ below about 400 GeV
which will be covered by the LHC search for $H/A\to\tau\tau$ with the
data that have already been recorded and are being analyzed. Such small values of $m_A$ are in some tension with the LHC Higgs coupling measurements
which currently set a lower bound on $m_A$ around 350 GeV.

We also pointed out that new LUX constraints on the SD scattering cross-section that have just been presented at the Moriond conference rule out
the well-tempered neutralino mass at least below 250 GeV for almost any value of $\tan\beta$.

Our findings indicate that the well-tempered bino-higgsino scenario
prefers split SUSY realization of the MSSM unless heavy Higgs doublets
are around the corner.
On the other hand, a well-tempered bino-higgsino may be consistent with
light stops in NMSSM for which low values of $\tan\beta$ are preferred
by the measured Higgs mass.
Moreover, the presence of the singlet in NMSSM provides important
modifications to the SI blind spot condition \cite{BS_NMSSM} and
substantially relaxes the constraints on $m_A$ from the LHC $\tau\tau$
searches as well as from the Higgs coupling measurements
\cite{alignment,tthsinglet,nmssmmixing}, hence allowing for large $H$
contribution to the SI scattering amplitude. Since in both, split SUSY
and NMSSM, the SI scattering cross-section may be below the irreducible
neutrino background the ultimate test of the well-tempered bino-higgsino
will be provided by searches for SD interactions between the LSP and nucleons.
In this kind of search LZ will cover the entire parameter space of the
well-tempered bino-higgsino. 

{\bf Note Added:}
During completion of this work Ref.~\cite{Huang:2017kdh} appeared which partially overlaps with some of the results of the present article.

\section*{Acknowledgements}
MB thanks Simon Knapen for interesting discussions.
This work has been partially supported by National Science Centre,
Poland, under research grants DEC-2014/15/B/ST2/02157,
DEC-2015/18/M/ST2/00054 and DEC-2012/04/A/ST2/00099, 
by the Office of High Energy Physics of the U.S. Department of Energy
under Contract DE-AC02-05CH11231, and by the National Science Foundation
under grant PHY-1316783. MB acknowledges support from the Polish 
Ministry of Science and Higher Education through its programme Mobility Plus (decision no.\ 1266/MOB/IV/2015/0).
PS acknowledges support from National Science Centre, Poland, 
grant DEC-2015/19/N/ST2/01697.

\appendix

\section{Criterion for resonant annihilation}
\label{App:kryterium}

The presence of relatively light $H$ and $A$ may affect the value of
$\Omega h^2$ very strongly if we are close to the resonance
$m_{\rm LSP}\approx m_H/2\,\, (m_A/2)$. The contribution to the LSP annihilation
cross-section from the $H$ and $A$ exchange decreases when we move away
from the resonance. However, it decreases smoothly and may be non-negligible
for a very wide range of the LSP masses. So, the LSP with relic abundance
dominated by a resonant annihilation goes smoothly into a well-tempered LSP.
In this Appendix we propose a simple criterion to define a border between
the well-tempered and resonant regions.

A given LSP should no longer
be considered as well-tempered if two conditions are fulfilled simultaneously.
Firstly, the mass of the particle exchanged in the s channel ($H$ or $A$
in the present case) is close to $m_{\rm LSP}/2$. Secondly, the contribution
from the considered s channel exchange influences the LSP relic abundance
in a substantial way. In order to quantify the above (qualitative) conditions
we calculate ratios of the LSP relic abundance using some limits. 
For the first condition we calculate $\Omega$ using two forms of the
propagator of the exchange particle: the actual one and the asymptotic
form valid far away from the resonance (more details below).
For the second condition we compare actual $\Omega$ with that obtained 
for decoupled $H/A$.

The method described below may be found in~\cite{DuchGrzadkowski}. Let us consider a dark matter particle (with mass $m$) annihilating via $s$ channel exchange of a particle with mass $M$ and total decay width $\Gamma$: 
\begin{equation}
\label{Appeq:sigmav_scalar}
\sigma v = \frac{\alpha}{(s-M^2)^2+\Gamma^2M^2}\,.
\end{equation}
For simplicity we assume $\alpha=\rm const$, which is generally not the case, however we are mainly focused on the effect on $\Omega h^2$ coming from the denominator in~\eqref{Appeq:sigmav_scalar}. Using dimensionless quantities $\delta\equiv 4m^2/M^2-1$, $\gamma\equiv\Gamma/M$ and considering non-relativistic approximation $s=4m^2/(1-v^2/4)\approx 4m^2(1+v^2/4)$ we get:
\begin{equation}
\label{Appeq:sigmav_scalar_app}
\sigma v = \frac{\alpha/M^4}{(\delta+v^2/4)^2+\gamma^2}\,.
\end{equation}
Let us now define $Y(x)\equiv\frac{n}{s}$, where $x=m/T$, and write
\begin{equation}
\frac{1}{Y(\infty)} - \frac{1}{Y(x_d)}=
m\,M_{\rm Pl}\,\frac{g_s}{\sqrt{g}}\sqrt{\frac{\pi}{45}}
\int_{x_d}^\infty \frac{\langle\sigma v\rangle}{x^2}\;{\rm d}x\,.
\end{equation}
Parameter $x_d$ is defined as a moment in thermal evolution of DM when the term $1/Y(x_d)$ starts to be small and can be safely neglected (we take $x_d=20$ however different values may influence slightly the results, especially for wide resonance). Dark matter relic abundance can be then calculated by double integration over $v$ and $x$:
\begin{align}
\label{Appeq:omega_res}
\Omega h^2 &= \frac{2.82\cdot 10^8}{\rm GeV}m\,Y(\infty)\\
&\approx
\frac{2.82\cdot 10^8}{\rm GeV}\frac{1}{M_{\rm Pl}}
\frac{\sqrt{g}}{g_s}\sqrt{\frac{45}{\pi}}
\left[
\frac{1}{2\sqrt{\pi}}
\int_0^\infty(\sigma v)v^2\;{\rm d}v
\int_{x_d}^\infty\frac{e^{-v^2x/4}}{\sqrt{x}}\;{\rm d}x
\right]^{-1}.
\end{align}
Note that we changed the usual order of integration. We will now perform the simpler integral over $x$, obtaining:
\begin{equation}
\frac{1}{2\sqrt{\pi}}
\int_0^\infty(\sigma v)v^2\;{\rm d}v
\int_{x_d}^\infty\frac{e^{-v^2x/4}}{\sqrt{x}}\;{\rm d}x=
\int_0^\infty(\sigma v)v\,{\rm erfc}(v\sqrt{x_d}/2)\;{\rm d}v\,.
\end{equation}
Now we can estimate the effect coming from the exchange of the resonant particle by taking the following ratio
\begin{equation}
\label{Appeq:ratio_eta}
\eta=
\frac{\Omega_{\rm no\;res.}}{\Omega_{\rm res.}}=
\int_0^\infty
\frac{v\,{\rm erfc}(v\sqrt{x_d}/2)}{(\delta+v^2/4)^2+\gamma^2}\;{\rm d}v
\left(
{\int_0^\infty 16m^4(\sigma v)v\,{\rm erfc}(v\sqrt{x_d}/2)\;{\rm d}v}\right)^{-1}
\,,
\end{equation}
where $\Omega_{\rm res.}$ refers to~\eqref{Appeq:omega_res} while
$\Omega_{\rm no\;res.}$ to $\sigma v$ being $\frac{\alpha}{M^4}$ for
$m<\frac{M}{2}$ or $\frac{\alpha}{s^2}$ for $m>\frac{M}{2}$.
As we already mentioned, there is no sharp border between the well-tempered
and resonant regions. So, there is no
one well defined value of $\eta$
from which the resonance should be considered to be important.
In the present work (see red dashed-dotted and dashed lines in
Figs.~\ref{fig:fixedMA} and \ref{fig:MAtanbeta}) we used $\eta=5,10$
to differentiate the well-tempered scenario from a resonant annihilation
via $H/A$ exchange.


\end{document}